\documentclass[showpacs,twocolumn,aps,prb,superscriptaddress,10pt]{revtex4-1}
\usepackage{amsmath,amssymb,amsfonts,bm,graphicx}
\usepackage{braket}

\usepackage{graphicx}
\usepackage{amsmath}
\usepackage{bm}
\usepackage{hyperref}
\hypersetup{%
	pdftitle={Competing order dynamics}, %
	pdfauthor={Zhiyuan Sun},%
	pdfpagemode={UseNone},%
	pdfstartview={FitH},%
	breaklinks=true,%
	citecolor=blue,%
	colorlinks=true,%
	linkcolor=blue,%
	urlcolor=blue
}

\usepackage[T1]{fontenc}
\usepackage{fourier}
\usepackage{baskervald}

\newcommand{\unit}[1]{\,\mathrm{#1}} 
\newcommand{\equa}[1]{Eq.~\eqref{#1}} %

\newcommand{\RomanNumeralCaps}[1]
{\textit{\MakeUppercase{\romannumeral #1}}}

\graphicspath{{figures/}}

\begin{document}

\title{Pump induced motion of an interface between competing orders}

\author{Zhiyuan Sun}
\affiliation{Department of Physics, Columbia University,
	538 West 120th Street, New York, New York 10027, USA}

\author{Andrew J. Millis}
\affiliation{Department of Physics, Columbia University,
	538 West 120th Street, New York, New York 10027, USA}
\affiliation{Center for Computational Quantum Physics, The Flatiron Institute, 162 5th Avenue, New York, New York 10010, USA}
\date{\today}

\begin{abstract}
We study the motion of an interface separating  two regions with different electronic orders following a short duration pump that drives the system out of equilibrium. Using a generalized Ginzburg-Landau approach and assuming that the main effect of the nonequilibrium drive is to transiently heat the system we address the question of the direction of interface motion; in other words, which ordered region expands and which contracts after the pump. 
Our analysis includes the effects of  differences in free energy landscape and in order
parameter dynamics and identifies circumstances in which the drive may act to increase the volume associated
with the subdominant order, for example when the subdominant order has a second order free energy
landscape while the dominant order has a first order one. \end{abstract}

\maketitle

\section{Introduction}
The control of electronic order via nonequilibrium drive is a problem of fundamental importance\cite{Domb.1983} and great current interest \cite{Zhang2016,McLeod2019,Cremin2019,Fausti2011,Kogar2019,Sun2019,Zhang2018,Gerasimenko2019,Nova.2019}. One recently studied situation is the case of competing orders, where  a nonequilibrium drive may suppress the order which is dominant in equilibrium, potentially allowing a different order, not observed in equilibrium, to arise.  This process may be viewed as a kind of ``order parameter steering'', in which the state of a system may be moved across a generalized free energy landscape by application of appropriate drive fields.

Recent experiments \cite{Zhang2016,McLeod2019,Cremin2019,Fausti2011,Kogar2019,Zhang2018,Gerasimenko2019} and theoretical analyses \cite{Kung2013,RossTagaras2019,Sun2019,Dolgirev2019} have addressed the situation in which a nonequilibrium drive pulse transiently drives the entire system into a disordered phase. After the drive ceases, order will re-form and depending on the dynamics an intermediate time regime may occur in which the system evolves to the local free energy minimum corresponding to a  subdominant metastable phase \cite{Sun2019}, before the system finally fully equilibrates.   However, many materials with competing electronic phases are characterized by phase coexistence, in which  different regions of a given sample are in different electronic phases separated by  relatively sharply defined interfaces. This situation may arise either from spatial inhomogeneity (quenched disorder) in the underlying material such that different spatial regions favor different orders, or as a dynamical effect arising from random initial conditions and domain wall pinning.  

Recent experiments \cite{Zhang2016,McLeod2019} have sharpened the physics questions. The system studied is a thin film of La$_{0.66}$Ca$_{0.33}$MnO$_3$; this material is a member of a class of systems that have two possible ground states: a ferromagnetic metal and a ``charge ordered", non-ferromagnetic insulator. The free energy differences between the phases are often very small, because modest (few Tesla) magnetic fields may switch the material from one state to another even at very low temperature\cite{Tokura06}.   In equilibrium, the La$_{0.66}$Ca$_{0.33}$MnO$_3$ film studied in Refs.~\cite{Zhang2016,McLeod2019}  exhibits a broad  transition at about $175K$ from a high temperature bad metal phase to a low temperature strongly insulating phase. The rapid change of resistance  across the transition region strongly suggests that the transition is first order with inhomogeneous broadening,   but a second order transition with a very rapid gap opening is not ruled out.  Cooling the system in a few-T magnetic field on the other hand produces a ferromagnetic metal state. Nano-optical measurements \cite{Zhang2016} indicate that the low T zero field cooled state is uniformly in the insulating phase.  A single pump pulse of moderate fluence at a frequency of $\sim 1.55 \unit{eV}$  produces small regions of ferromagnetic metal, which survive over timescales of hours to days if the temperature is kept below $T_{\text{spinodal}}\approx 120 \unit{K}$. Subsequent pulses of optical excitation with similar fluence are found not to create new domains but rather to expand the existing metallic domains, with the domains remaining in the expanded size over long times provided the temperature is kept low enough. After a sufficient number of pulses the entire sample is fully transformed to an apparently homogeneous metallic state. 

Both the physics of  manganite materials and specifics of the experiment are complicated. In manganites generally and very probably in the film studied in Refs.~\cite{Zhang2016,McLeod2019} the energetics of strain fields produced by the charge ordering will be important to  the  dynamics \cite{Tokura06,Seman12}, while the response to the initial pump pulse shows that the charge ordered state is weaker  in some areas of the sample than in others. But the experiments raise general and  fundamental questions that are independent of the specifics of the particular experimental system studied here: why does the nonequilibrium drive act to expand one phase at the expense of the other, and under what circumstances may the expanded phase be the one disfavored in equilibrium. These questions are the main focus of this paper. 

The issue of the expansion of one phase with respect to another is appropriately addressed via an order parameter theory, which we take to have relaxational dynamics and generalized forces arising from functional differentiation of an energy-like function of the order parameters.  The length ($\gtrsim \unit{nm}$) and time ($\gtrsim \unit{ps}$) scales relevant to interface motion mean that the details of the pump pulse and other aspects of microscopic dynamics are not important: we can simply view the pump as providing a time dependence for the coefficients  of the order parameter theory.   In equilibrium, the energy-like function will just be the equilibrium free energy landscape, which will be characterized by two locally stable extrema, corresponding to the two competing states. The configuration of a spatially uniform system will be concentrated at the extremum with the lower free energy.  A pump pulse will change the landscape, thus producing generalized forces that will drive the system away from the equilibrium extremum. For strong pump pulses the order parameters will be driven to zero and one must then consider the dynamical reformation of the ordered state from a fully disordered configuration \cite{Sun2019}. However, for weaker pump pulses the order parameter for a homogeneous system (or an inhomogeneous system far from an interface) will be only moderately perturbed; the system will remain  in the basin of the attraction of the equilibrium free energy minimum and  will simply relax back to the starting configuration when the pump-induced changes in energy parameters decay away. However,  at or near an interface between two orders the value of each order parameter will be far from the extrema and the dynamics need not be a simple attraction to a fixed point.  

To analyse this dynamics we make a time-dependent mean field approximation, so the theory is a set of nonlinear deterministic  partial differential equations for order parameter fields, with an initial condition including the presence of the interface and dynamics arising from transient pump-induced modifications of the energy parameters. The pulse will transiently reduce the amplitude of each order parameter and the post-pulse dynamics will involve evolution of the two order parameters, which will both grow and compete with each other. We expect that the long time limit  will be a new steady state in which the interface have moved to a new position. Our aim is to determine the direction of motion of the interface. 

Interface motion arises from an asymmetry between order parameters, which in turn may have several origins, including a difference in relaxational time constants (this was the focus of our previous work \cite{Sun2019} on bulk phases), a difference in generalized forces arising from different structures of the free energy landscapes of the different orders, and a difference in coupling to the pump. We consider all three cases. A key result is that the difference in structure of the free energy landscape arising in the case of competing phases with first and second order free energy landscapes respectively causes the interface to move in order to expand the phase with the second order transition, even if this phase is not the global free energy minimum. 

One important issue requires discussion. Unless a system is tuned exactly to the degeneracy point between two uniform orders, an interface can be stabilized only by pinning due to quenched randomness in the underlying material. Interface dynamics in the presence of pinning is subtle and complicated, but not directly relevant to the experiments of interest, which show that interfaces may be stabilized at essentially arbitrary positions. We take the view that pinning acts on long length scales and for large order parameter amplitudes, so we employ a model without explicit pinning to determine the direction of motion of the interface. 

The rest of this paper is organized as follows. In Section~\ref{sec:formalism} we present our formalism for order parameter dynamics,  introduce the  energy landscape we study,   construct the static interface that is the starting point of our calculations and give the specific pump profiles we use.  Section~\ref{sec:linear_response}  presents the case of a weak, short duration pump, where analytical results can be obtained. Section~\ref{sec:numerical} presents  numerical results  beyond linear response and provides qualitative understanding of them, treating the different sources of asymmetry between the phases. Section~\ref{sec:discussion} is a summary and outlook.

\begin{figure}
	\includegraphics[width=1.0 \linewidth]{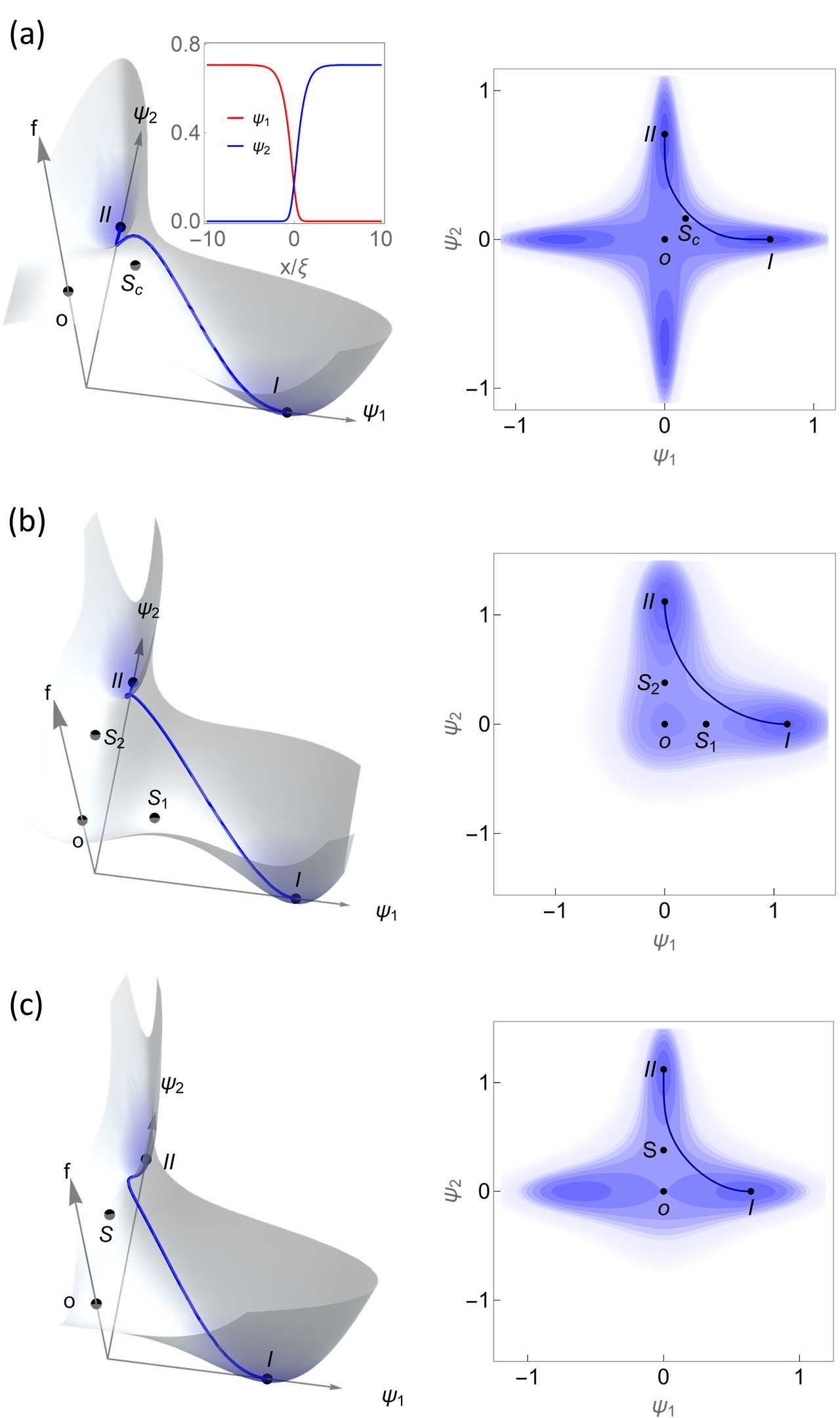}
	\caption{Left/right panel is the three dimensional representation/contour plot of the free energy landscape of the competing order systems, with lower energy appearing bluer. Blue lines are the projections of the order parameter profiles across the interfaces. (a) Even order terms up to $\psi^4$ lead to second order landscape. $S_c$ is the saddle point in the diagonal direction. Inset is the order parameter spatial profile across the interface between the two regions for $\alpha_i=-1$ and $c=100$. (b) The  term $\lambda_i \psi_i^3$ leads to first order landscape and the point $O$ ($\psi_1=\psi_2=0$) is also a local minimum for $\alpha_i>0$. $S_{1}$/$S_{2}$ are saddle points on the $\psi_2=0$/$\psi_1=0$ axes. (c) $\psi_1$ has second order landscape while $\psi_2$ has first order landscape. $S$ and $O$ are saddle points. }
	\label{fig:landscape}
\end{figure}

\section{Statics: free energy and interface\label{sec:formalism}}
\subsection{General formulae}
We consider a system in which the important degrees of freedom are two space-time dependent order parameter fields $\psi_{i=1,2}(\mathbf{r},t)$ obtained from a fundamental theory by integrating out microscopic degrees of freedom. Applied to manganites\cite{Zhang2016,McLeod2019}, $\psi_2$/$\psi_1$ could be associated with the antiferromagnetic insulator/ferromagnetic metal phase. The order parameter fields evolve according to dissipative (relaxational or time-dependent-Ginzburg-Landau ``TDGL'' or ``Model A'') dynamics \cite{Gorkov1968, Cyrot1973, Hohenberg1977,Lemonik.2017,Sun2019}  \begin{align}
\frac{1}{\gamma_i }\partial_t \psi_i(\mathbf{r},t)
&=\mathcal{F}_i\equiv
-\frac{\delta F(t)}{\delta \psi_i (\mathbf{r},t)} 
\,.
\label{eqn:TDGL}
\end{align}
Here the $\gamma_i$ are time constants and the generalized forces $\mathcal{F}_i$ are obtained from an energy functional $F$ defined as an integral over an energy density of the general form
\begin{equation}
F[\psi_1,\psi_2;t] = \int d^D\mathbf{r} \, \left(f_1[\psi_1;t] + f_2[\psi_2;t] +f_c[\psi_1,\psi_2] \right)
\label{Fdef}
\end{equation}
where in equilibrium the  two free energies $f_{1,2}$ of the individual orders have locally stable extrema at some nonzero values of $\psi_{1,2}$ respectively and $f_c$ expresses the physics that presence of $\psi_1$ suppresses $\psi_2$ and conversely such that the only stable extrema have at least one of $\psi_1,\psi_2=0$. 
The  free energy density of order $i=1,2$ is assumed to be of the general form
\begin{align}
& f_i= \alpha_i(t) \psi_i^2 + \lambda_i \psi_i^3  + \psi_i^4  + \left( \xi_{i0} \nabla \psi_i \right)^2\,
\,
\label{eqn:free_energy}
\end{align}
and the competing term is 
\begin{equation}
f_c =c \psi_1^2 \psi_2^2
\,.
\label{eqn:fc}
\end{equation}
Note that to describe a first order energy landscape we used a free energy with 
a cubic term. Alternative implementations of first order energy landscapes that 
maintain a $Z_2$ symmetry may be written, but give the same qualitative behavior (in particular the same mean field dynamics) as does the free energy we have written.

Without loss of generality we normalize the fields and energies such that  $\psi_i$, $f_i$, $\alpha_i$ and $\lambda_i$ are all dimensionless, the coefficient of the quartic term is unity and $F$ is measured in units of a characteristic condensation energy density.  The term proportional to $c$ expresses the competition between the phases and in the cases of primary interest we expect $c$ to be large and positive. The difference between the two orders are contained in the dimensionless parameters $\alpha_i, \lambda_i$. In the following, we suppress the label $i$ whenever possible without loss of clarity. 

In equilibrium the individual  free energy functions may either have a second order ($\lambda=0$) or first order ($\lambda\neq 0$) structure.  In the second order case, the equilibrium ground state is ordered, with $\psi_m=\sqrt{-\alpha/2}$ if $\alpha <0$. If both $\alpha<0$ we have two possible ordered states. If  $c>2$ and $-\frac{c}{2} \alpha_1 > -\alpha_2 > -\alpha_1$ then  there is a global minimum corresponding to order \RomanNumeralCaps{2} and a metastable minimum corresponding to order \RomanNumeralCaps{1}, as shown in Fig.~\ref{fig:landscape}(a). Also as shown in this panel there is a least-energy path connecting the two minima along which both $\psi_{1}$ and $\psi_2\neq 0$. The least-energy path bypasses the origin, which is a local maximum.

If $\lambda_i\neq 0$ the transition for  $\psi_i$  is first order. Taking for concreteness the   case of $\lambda <0$, a state with the order parameter
\begin{align}
\psi_{m} =\left( -3\lambda + \sqrt{9\lambda^2 - 32 \alpha} \right)/8
\end{align}
becomes locally stable if $\alpha < 9\lambda^2/32$, and becomes the global ground state if $\alpha < \lambda^2/4$. If $\alpha>0$ (i.e., if the system is within the lower spinodal) the origin is a local minimum and is separated from the global minimum by an intermediate saddle point at $\psi_s=\left( -3\lambda - \sqrt{9\lambda^2 - 32 \alpha} \right)/8$. The saddle point along $\psi_i$ direction is labeled by $S_i$ in Fig.~\ref{fig:landscape}(b).

If both orders are below the transition and within the spinodal regions, the two orders are separated by a least energy path, as shown by Fig.~\ref{fig:landscape}(b).  Finally, Fig.~\ref{fig:landscape}(c) shows the case where one of the transition is first order and one is second order. In this case the least energy path takes a highly asymmetric trajectory in order parameter space. 

\subsection{Domain Wall}
We construct a domain wall between the two regions by numerically minimizing $f$ subject to the boundary conditions $\psi_1(x\rightarrow -\infty)=\psi_{1m}$, $\psi_1(x\rightarrow \infty)=0$ and  $\psi_2(x\rightarrow -\infty)=0$, $\psi_2(x\rightarrow \infty)=\psi_{2m}$.
In the second order-second order and first order-first order cases a good approximation to the domain wall profile is
\begin{equation}
\psi_i = \psi_{im} \left( \pm \tanh\left(\frac{x\mp \delta_c}{\xi} \right) +1 \right)/2
\,
\label{eqn:interface_profile}
\end{equation}
where $+/-$ corresponds to $i=2/1$ and the difference between the coherence lengths has been neglected. 
In the case of large $c$, the two phases strongly repel each other such that $\delta_c$ is very large, the order parameter trajectory passes near the saddle point $\psi_1\sim \psi_2\sim 1/\sqrt{c}$ but the length scales $\xi\sim \xi_0/\sqrt{\alpha}$ remain set by the coherence length, leading to the domain wall structure shown by the inset of Fig.~\ref{fig:landscape}(a). 

\subsection{Time dependence of landscape parameters}

\begin{figure}[htbp]
	\includegraphics[width= 0.9 \linewidth]{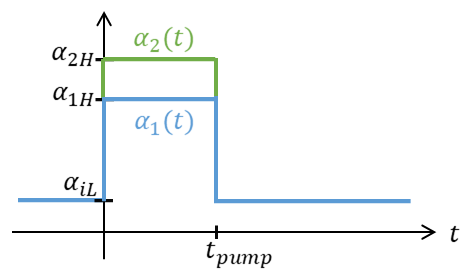}
	\caption{The time dependence of $\alpha_i$ across the pumping process.}
	\label{fig:Quench_shape}
\end{figure}

Motivated by the idea that the main effect of the pump is to transiently heat the system we have assumed that the main time dependence is in the quadratic coefficients $\alpha_{i}(t)=\alpha_i+a_i(t)$ (which would carry the main temperature dependence in the equilibrium Ginzburg-Landau approach).  Representative time dependences are shown in Fig.~\ref{fig:Quench_shape} where the quadratic coefficients change from their static values $\alpha_{iL}$ to higher values $\alpha_{iH}$ during the time pump is on.

If the two free energy minima have different energies, then in the absence of pinning the interface will move so as to expand the size of the region with the lower energy minimum. In practice domain walls may become pinned by impurities on experimental timescales\cite{Zhang2016,McLeod2019}. To effectively include the effect of impurity pining, we set the energies of the two minima the same. We will return to this point in the conclusions.



\section{Interface motion: linear response to pump \label{sec:linear_response}}
In this section, we assume the pump is of sufficiently small amplitude and short duration that it is enough to consider the linear response of the order parameter configuration $\psi_{i}=\psi_i(x) + \phi_{i}(x,t)$ to the pump $a_i(t)$. The TDGL equation \equa{eqn:TDGL} linearized around the interface solution $\psi_i(x)$ reads
\begin{equation}
\frac{1}{\gamma_i} \partial_t \phi_i
=-  \sum_j \frac{\delta^2 F}{\delta \psi_i \delta \psi_j} \phi_j - 2  a_i \psi_i
\,
\label{eqn:TDGL_linearized}
\end{equation}
where the second order functional derivative is the quadratic kernel for the free energy cost due to the small fluctuation $\phi(x)$ around the interface configuration $\psi(x)$. Define the `re-scaled' field $\phi_i^\prime=\phi_i/\sqrt{\gamma_i}$ which transforms \equa{eqn:TDGL_linearized} to
\begin{equation}
\partial_t \phi_i^\prime
=-  \sqrt{\gamma_i} \sum_j \frac{\delta^2 F}{\delta \psi_i \delta \psi_j} \sqrt{\gamma_j} \phi_j^\prime - 2  \sqrt{\gamma_i} a_i \psi_i
\,.
\label{eqn:TDGL_linearized2}
\end{equation}
The coefficient of the first term on the right hand side can be viewed as a Hermitian linear operator $\hat{L}$ acting on the two component field $\left(\phi_1^\prime(x),\phi_2^\prime(x) \right)$. Decomposing $\phi^\prime$ using normalized eigenfunctions $f_n(x)$ of $\hat{L}$ with eigenvalue $\lambda_n$: $\phi^\prime(x,t)=\sum_n c_n(t) f_n(x)$, inserting it into \equa{eqn:TDGL_linearized2} and taking the inner product with $f_n(x)$ results in an equation for the time-dependent expansion coefficients
\begin{equation}
\partial_t c_n
= - \lambda_n c_n - 2\langle f_n| \sqrt{\hat{\gamma}} \hat{a}(t) |\psi\rangle
\,
\label{eqn:linear_coeficient}
\end{equation}
with the solution
\begin{equation}
c_n(t)=\int_{-\infty}^tdt^\prime e^{-\lambda_n(t-t^\prime)}
2\langle f_n|\sqrt{\hat{\gamma}} \hat{a}(t^\prime) |\psi\rangle
\,.
\label{eqn:linear_coeficient_soln}
\end{equation}

Since $\hat{L}$ is positive semi-definite, we have $\lambda_n \geq 0$. For the components with $\lambda_n >0$, the solution $c_n(t)$ to \equa{eqn:linear_coeficient} vanishes at long times after $\hat{a}(t)\rightarrow 0$, so  within linear response these components don't contribute to any change of the $\psi$ configuration, thus lead to no interface motion.
However, there is one eigenfunction with exactly zero eigenvalue: $f_0(x)=\frac{1}{d_0} \sqrt{\hat{\gamma}^{-1}} \partial_x \psi$ where $d_0 =\sqrt{\langle \partial_x \psi |\hat{\gamma}^{-1}|\partial_x \psi \rangle}$ is the normalization factor, corresponding to an infinitesimal translation of the interface which does not change the free energy. For this zero mode, \equa{eqn:linear_coeficient} yields 
\begin{align}
c_0
&= -2\int dt 
\langle f_0| \sqrt{\hat{\gamma}} \hat{a}(t) |\psi \rangle
\notag\\
&=- \frac{1}{ d_0} \int dt
\left( -a_1(t) \psi_{m1}^2 + a_2(t) \psi_{m2}^2 \right)
\,.
\label{eqn:linear_zero_coeficient}
\end{align}
where $\psi_{im}$ is the order parameter value far away from the interface.

Using the relation between the zero eigenfunction $f_0$ and the interface translation $\Delta x$  we obtain
\begin{align}
\Delta x &=\frac{1}{ d_0^2} \int dt
\left( -\psi_{m1}^2 a_1(t) + \psi_{m2}^2 a_2(t) \right)
\label{eqn:linear_motion}
\end{align}
where 
\begin{align}
d_0^2 = \int dx \left( \frac{\left(\partial_x \psi_1 \right)^2}{\gamma_1} + 
\frac{\left(\partial_x \psi_2 \right)^2}{\gamma_2} \right)
\approx \left( \frac{|\alpha_1|}{\gamma_1 \xi_1} +  \frac{|\alpha_2|}{\gamma_2 \xi_2} \right)
\,.
\end{align}

We now interpret Eq.~\ref{eqn:linear_motion}. The integrand is in effect a force pushing the interface to move. The amount of motion is linearly proportional to the pump fluence and the coherence length. The direction of motion is determined by the effect of the pump on each order and the properties of each order (encoded in the values $\psi_{mi}$ of the order parameters far from the interface). Note that dynamics does not directly enter: if the only asymmetry is between the relaxation rates, the interface does not move to linear order in the pump fluence. We shall see that at higher orders in pump fluence, the relaxation rate is also important. 

Because the asymptotic value $\psi_m$ extremizes the free energy, to linear order in the pump field $a \psi_m^2$ is just the pump induced change in free energy, so we find that to this order the pump acts to move the interface so as to expand the order which is transiently favored by the pump.  Beyond linear response the physics may be different, particularly in the case of the first order energy landscape where the energy also involves the parameter $\lambda$.

\section{Interface motion: numerical results \label{sec:numerical}}
\subsection{Overview} In this section we present results obtained from numerical solutions of the equations presented in Section~\ref{sec:formalism}.  We consider representative examples of the three general cases: second order-second order, second order-first order and first order-first order, investigating different combinations of relaxation rates and drives. In each case we numerically construct the domain wall solution and then consider its dynamical evolution, focussing on which order expands and which contracts.  For simplicity we take the $\alpha_i$ to have the step function shape  shown in Fig.~\ref{fig:Quench_shape}.  

\subsection{Second order-second order}
\begin{figure}
	\includegraphics[width= 0.9 \linewidth]{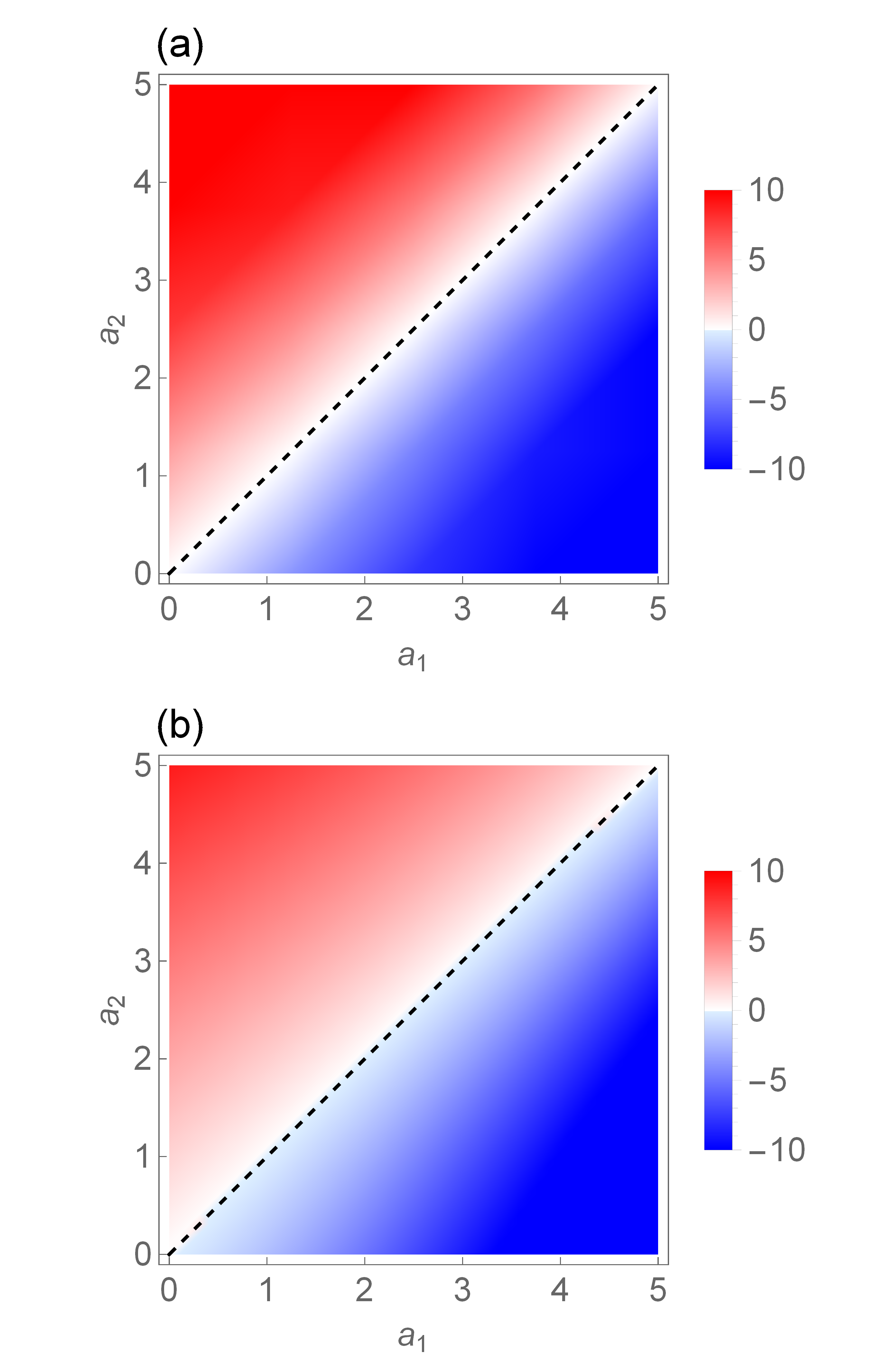}
	\caption{False color representation  of interface motion as a function of  drive difference $a_i=\alpha_{iH}-\alpha_{iL}$ when both orders have second order landscapes.  Red means that the interface moves to expand region $\RomanNumeralCaps{1}$ and blue means that the interface moves to expand region $\RomanNumeralCaps{2}$. The dashed line is the line of zero motion as predicted by linear response theory \equa{eqn:linear_motion}. The parameters used are $\alpha_{iL}=-1$, $c=4$, $\xi_0=1$, $t_{\text{pump}}=1$. Panel (a) has $\gamma_1=\gamma_2=1$ and panel (b) has $\gamma_1=0.5$, $\gamma_2=1$. }
	\label{fig:interface_motion_22}
\end{figure}

\begin{figure}
	\includegraphics[width= 0.8 \linewidth]{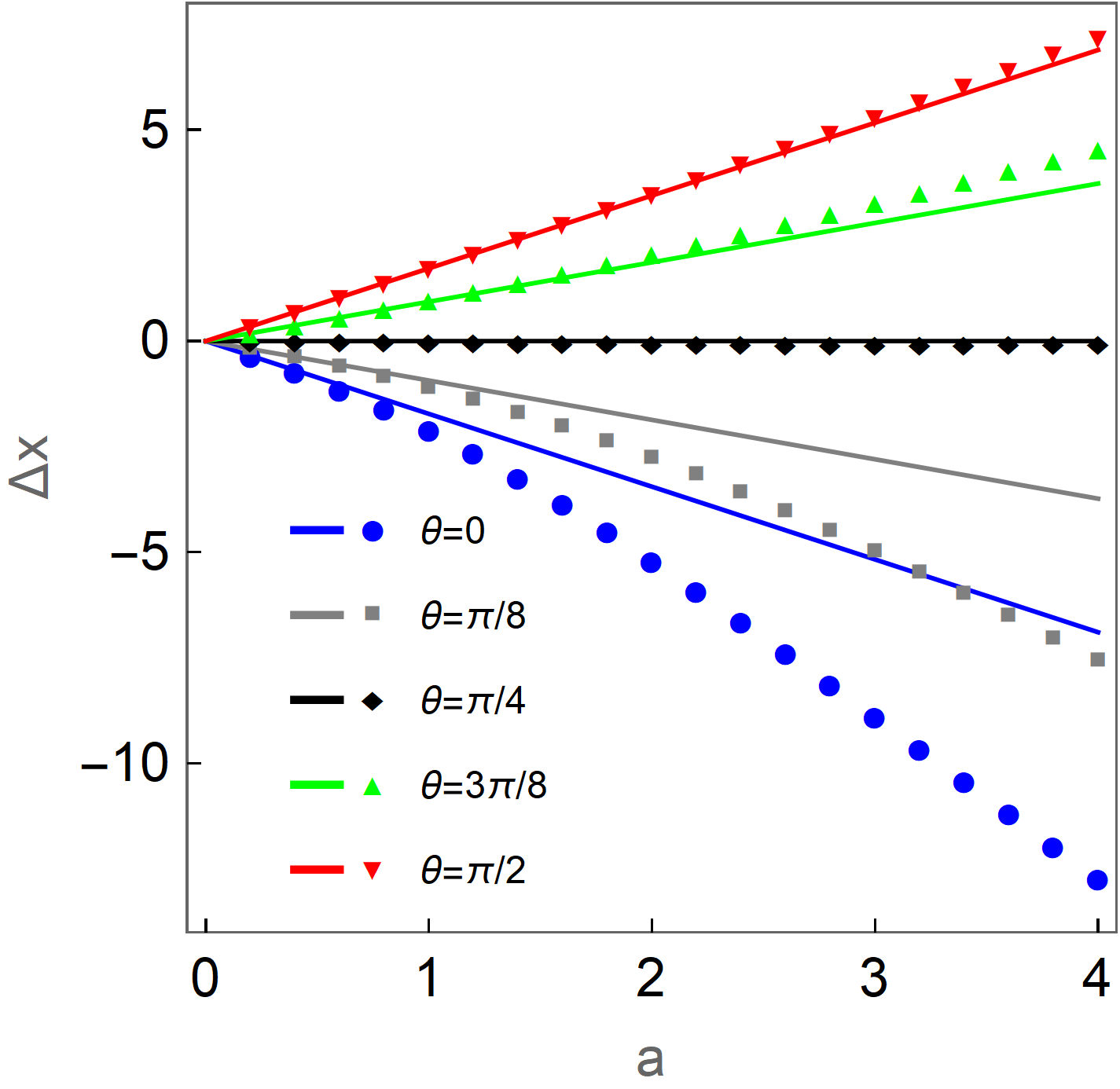}
	\caption{The interface motion as a function of the pump strength $(a_1,\,a_2)=a(\cos\theta,\, \sin \theta)$ in the case of second order v.s. second order landscapes. The dots are from numerically exact results in Fig.~\ref{fig:interface_motion_22}(b) while the solid lines are the linear response predictions from \equa{eqn:linear_motion} with the same parameters. Different angles $\theta$ correspond to different radial directions in Fig.~\ref{fig:interface_motion_22}(b), e.g., $\theta=\pi/4$ means looking along the dashed line. 
	 }
	\label{fig:linear_nonlinear_22}
\end{figure}

In this subsection we consider the motion of an interface separating two order parameters, each of which  in isolation has a second order energy landscape (Fig.~\ref{fig:landscape}(a)). Representative results are shown in Fig.~\ref{fig:interface_motion_22}. We find that (as seen in the linear response calculation of Section \ref{sec:linear_response}) the interface moves so as to expand the  phase which is less strongly affected by the pump and that this conclusion holds even for a substantial difference in relaxation rates (cf position of dashed line in Fig.~\ref{fig:interface_motion_22} (b)). This is in contrast to the case in which the pump fully destroys both orders, where the long time state is strongly affected by differences in dynamics \cite{Sun2019}. In fact  the linear response result accurately captures the amplitude of interface motion even for rather large fluencies, see Fig.~\ref{fig:linear_nonlinear_22} for a comparison.

The physics behind the role played by the differential effect of the pump  may be seen from consideration of the limit in which  $\psi_2$ is strongly suppressed  by the pump while $\psi_1$ is barely affected. Order \RomanNumeralCaps{2} will need some time $t_r \sim |\alpha_{2H}/\alpha_{L}|t_{\text{pump}}$ to recover after the pump. Before $\psi_2$ fully recovers, the $\psi_1$ front tends to translate with the `soliton' solution
\begin{equation}
\psi_1(x,t) = \sqrt{\frac{|\alpha_i|}{2}} \frac{1}{2} \left( - \tanh((x-vt)/\xi) +1 \right)
\,
\label{eqn:translate}
\end{equation}
where $\xi=\xi_0 \sqrt{8/|\alpha_L|}$ and the velocity is $v= \frac{3}{2} |\alpha_L| \gamma_1 \xi$. Thus phase \RomanNumeralCaps{1} domain expands to the right as long as $\psi_2$ has not recovered enough to stop it. This translation continues for $t_r$ and thus the amount of domain expansion is
\begin{equation}
\Delta x \approx v t_r = \frac{3}{2} |\alpha_{2H}| \gamma_1 \xi
t_{\text{pump}}
\,.
\label{eqn:distance_expand}
\end{equation}

Further insight into the weak effects of a difference of relaxation rates may be obtained from consideration of the limit of very strong pump fluence $|\alpha_H t_p| \gg 1$. In this limit the dynamics during the pump is linear and is to a good approximation  
\begin{equation}
\psi_i(x,t) = \frac{1}{2} \sqrt{\frac{|\alpha_{L}|}{2}}
e^{-2\gamma_i t} \left( \pm \mathrm{Erf}(x/\sqrt{8\gamma_i t} ) +1 \right)
\,
\label{eqn:error_function}
\end{equation}
for $\gamma_i t \gg (\xi/\xi_0)^2$ where $\mathrm{Erf}$ is the error function. Thus independent of the relaxation times  the interface stays fixed during the pump, although the two orders are suppressed to different levels.  After the pump is turned off both orders  recover; the difference in recover rates essentially compensate for the difference in suppression, leading again to a very weak diependence of interface position on order parameter relaxation time scales. 


\subsection{First order-first order}
\label{sec:first_first}
\begin{figure}
	\includegraphics[width= \linewidth]{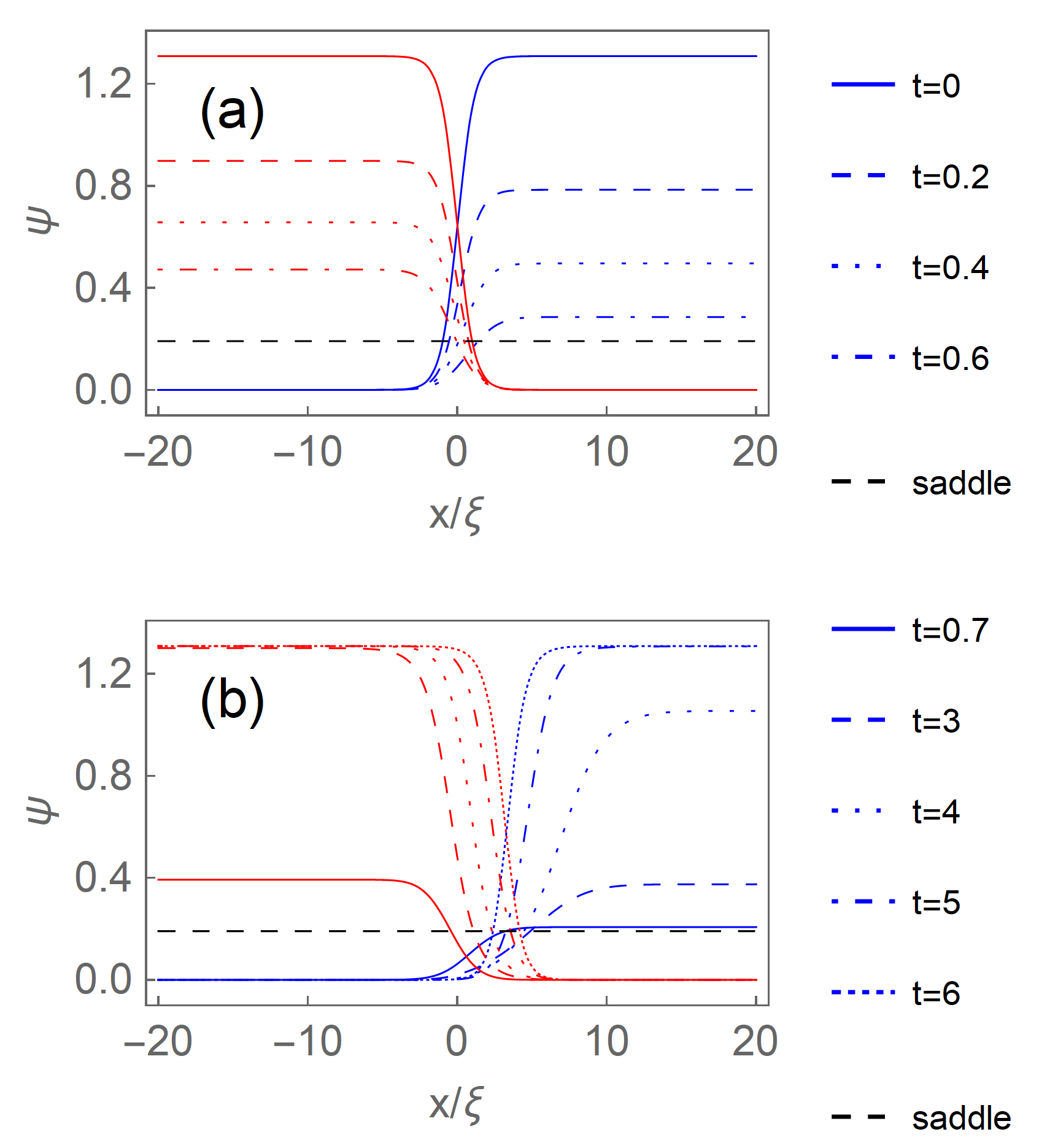}
	\caption{The (a) suppression and (b) recovery of the order parameters during and after the pump ($t_{\text{pump}}=0.7$) in the first order-first order case. Red curves are values of $\psi_1$ and blue ones are $\psi_2$. The parameters are $\alpha_{iL}=0.5$, $\alpha_{iH}=2.23$, $\lambda_i=-2$, $c=1.5$, $\xi_0=1$, $\gamma_1=0.7$, $\gamma_2=1$.}
	\label{fig:interface_suppress}
\end{figure}
\begin{figure}
	\includegraphics[width= 0.9 \linewidth]{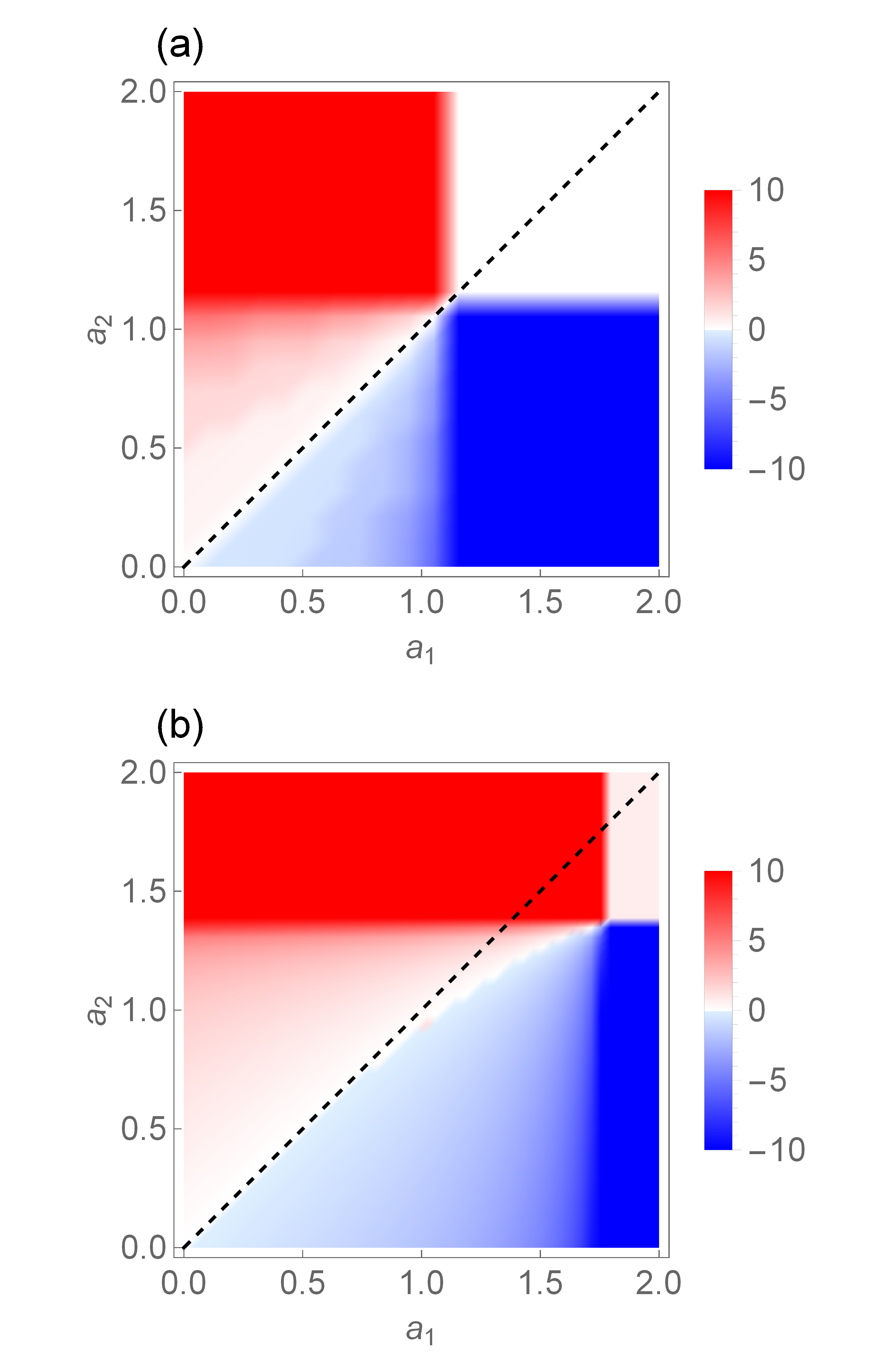}
	\caption{The amount of interface motion as a function of $a_i=\alpha_{iH}-\alpha_{iL}$ when both orders have first order landscapes. Red means that the interface moves to expand region $\RomanNumeralCaps{1}$ and blue means that the interface moves to expand region $\RomanNumeralCaps{2}$. The dashed line is the line of zero motion as predicted by linear response theory \equa{eqn:linear_motion}. The parameters used are $\alpha_{iL}=0.5$, $\lambda_i=-2$, $c=1.5$, $\xi_0=1$, $t_{pump}=1$. Panel (a) has $\gamma_1=\gamma_2=1$ and panel (b) has $\gamma_1=0.7$, $\gamma_2=1$.  }
	\label{fig:interface_motion_11}
\end{figure}

In this subsection we consider the motion of an interface separating two order parameters, each of which has in isolation a first order energy landscape (Fig.~\ref{fig:landscape}(b)). The interface profile is as shown in Fig.~\ref{fig:interface_suppress}.  Results of our calculations are shown in Fig.~\ref{fig:interface_motion_11}.  In broad terms the physics of the second order-second order situation applies also to the first order-first order case: if the pump couples more strongly to phase \RomanNumeralCaps{2}, the phase \RomanNumeralCaps{1} domain expands.

However, two important differences arise, related to differences in the structure of the relevant energy landscape. First, we observe that along the line $\psi_1=0$, the free energy curve passes over a local maximum (saddle, when variations in the $\psi_1$ direction are included. In the vicinity of the saddle, order parameter dynamics become slow. Thus if for example  order \RomanNumeralCaps{2} relaxes faster than order \RomanNumeralCaps{1}  ($\gamma_2 > \gamma_1$) then it may be  that under the action of the pump $\psi_2(x\rightarrow \infty)$ is driven close to  its saddle point $S_2$  while $\psi_1(x\rightarrow -\infty)$ is not suppressed enough to get close to its saddle point $S_1$.  In this case, when the pump is turned off the slow near saddle point dynamics means that  order \RomanNumeralCaps{2} will recover very slowly  (similar to critical slowing down) while $\psi_1$ will recover faster to its equilibrium value, after which the domain associated with  $\psi_1$ expands. This is numerically illustrated in Fig.~\ref{fig:interface_suppress} and the amount of interface motion is plotted in Fig.~\ref{fig:interface_motion_11}(b).

Further, in the first order situation if the temperature is within the spinodal region the origin may also  be locally stable.  If the pump is strong enough ($2 |\alpha_{2H}| \gamma_2 t_p > \ln (\psi_{2m}/\psi_{2s})$) to push $\psi_2$ beyond the saddle point $S_2$ into the basin of attraction of the origin, the entire phase \RomanNumeralCaps{2} domain may be trapped into the $\psi=0$ local minimum after the pump so that  phase \RomanNumeralCaps{1} domain expands until the whole system is depleted, as shown by the red regions in Fig.~\ref{fig:interface_motion_11}.

If the pump is even stronger such that both $\psi_2$ and $\psi_1$ are suppressed beyond their saddle points ($2 |\alpha_{iH}| \gamma_i t_p > \ln (\psi_{im}/\psi_{is})$), the whole system is trapped into the metastable  disordered phase $\psi_i=0$  as shown by the white regions in Fig.~\ref{fig:interface_motion_11}.

\subsection{Second order-first order}

\begin{figure}[t]
	\includegraphics[width= 0.8 \linewidth]{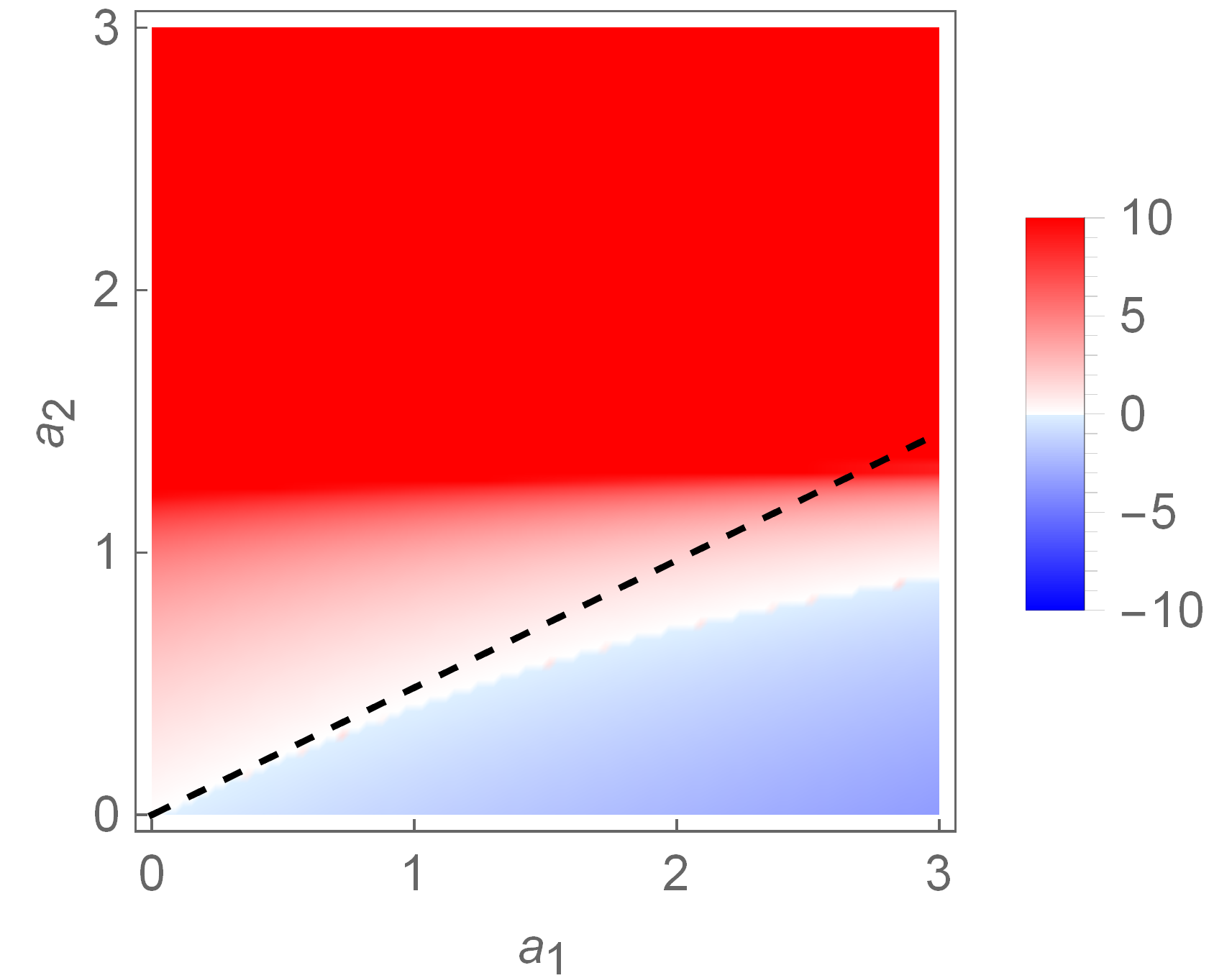}
	\caption{The amount of interface motion as a function of $a_i=\alpha_{iH}-\alpha_{iL}$ when $\psi_1$ has second order landscape while $\psi_2$ has first order. Red means that the interface moves to expand region $\RomanNumeralCaps{1}$ and blue means that the interface moves to expand region $\RomanNumeralCaps{2}$. The dashed line is the line for zero motion ($\Delta x=0$) as predicted by linear response theory \equa{eqn:linear_motion}. The difference of its slope relative to Fig.~\ref{fig:interface_motion_11} arises from the difference in asymptotic values of the $\psi_{im}$.
The parameters used are $\alpha_{2L}=0.5$, $\alpha_{1L}=-1.66$, $c=1.5$, $\lambda=-2$, $\xi_0=1$, $t_{pump}=1$ and $\gamma_1=\gamma_2=1$.}
	\label{fig:a1h_a2h_12}
\end{figure}

In this subsection we consider the motion of an interface separating two order parameters, one of which has in isolation a first order energy landscape and the other a second order landscape (Fig.~\ref{fig:landscape}(c)). 
In linear response, the motion of the interface depends only on which order is more strongly affected by the pump, but beyond linear response the  difference in  free energy landscape provides a natural asymmetry between the two side of the interface, tending to favor expansion of the phase with the second order landscape. Numerical results are presented in Fig.~\ref{fig:a1h_a2h_12}.

One way to understand this phenomenon is via the saddle point argument of the previous section. Even if the relaxation rates are the same and the pump simply raises the temperature such that both orders are weakened, the phase with the first order landscape  (here, \RomanNumeralCaps{2}) will be driven towards its saddle point,  which makes its recovery very slow or beyond, into the near origin region where it is attracted to zero (deep red  region of Fig.~\ref{fig:a1h_a2h_12}).  By contrast the second order landscape of the other phase means that it recovers quickly and then expands with a velocity on the order of $v\sim \gamma \xi \left(\sqrt{f_s}-\sqrt{f_m} \right) $ where $f_s=f_2(\psi_s)$ is the free energy at the saddle point and $f_m=f_1(\psi_m)$ is that at the phase $\RomanNumeralCaps{1}$ minimum. The amount of interface motion as a function of $\alpha_{iH}$ is shown in Fig.~\ref{fig:a1h_a2h_12}.

For stronger pump which suppresses $\psi_2$ beyond point S, $\psi_2$ will not recover after the pump is gone since there is a potential barrier in the direction of increasing $\psi_2$ (see Fig.~\ref{fig:landscape}(c)) while $\psi_1$ does not have this problem. After pump is gone, the entire phase $\RomanNumeralCaps{2}$ region will be suppressed to zero order and phase $\RomanNumeralCaps{1}$ region expands with the velocity $v= \frac{3}{2} |\alpha_{1L}| \gamma_1 \xi$ until the whole sample is transformed into
phase $\RomanNumeralCaps{1}$. This phenomenon happens robustly as long as the pump is strong enough, whether it prefers to affect phase $\RomanNumeralCaps{2}$ or not. Indeed, if one follows any straight line from the origin to large values of $a_i$, the interface motion is always to the right (represented by red color) in Fig.~\ref{fig:a1h_a2h_12}. Note that this phenomenon happens also for uniform samples without preformed phase $\RomanNumeralCaps{1}$ domains. In this case, any strong enough pump could destroy phase $\RomanNumeralCaps{2}$ to a disordered state, and random $\RomanNumeralCaps{1}$ domains would appear afterwards.

\section{Conclusion \label{sec:discussion}}
We have studied the possible mechanisms of pump induced motion of the  interface between two different electronic phases, making the assumptions that on the relevant timescales the pump in effect provides a transient change to the Ginzburg-Landau parameters that is qualitatively similar to a change in temperature, and that the relevant order parameter dynamics can be described by a time-dependent Ginsburg-Landau equation with relaxational dynamics (\equa{eqn:TDGL}).  
Relaxational dynamics is clearly valid near and above the transition temperature\cite{Lemonik.2017}. However, deep inside the ordered phase a term in the Ginzburg-Landau equation proportional to the second order derivative ($\partial_t^2$)  of the order parameter  is present and will lead to propagating modes describing order parameter fluctuations \cite{sun2020collective}. These fluctuations are not directly relevant to the physics we consider, which relates to large amplitude changes in the order parameter magnitude over wide areas although as an interface moves it may dissipate energy by emitting order parameter fluctuations. We further note that in the presence of a constant force $F$ the solution of
$\frac{1}{\omega_0^2} \partial_t^2 \psi + \frac{1}{\gamma} \partial_t \psi=F$ is $\psi=\gamma Ft$,
independent of the coefficient of the second order derivative. This solution applies also to a time-dependent force provided $\partial_tF \ll \omega_0^2/\gamma$. Thus in this limit, the $\partial_t^2$ term will affect initial transients and renormalize the dissipation but not change the qualitative physics. 

While the direction of interface motion can depend on many factors, we found that in general the most important issue was the differential effect of the pump on the different phases: the phase that is more strongly affected by the pump shrinks, while the phase that is less strongly affected grows. 
One expects on general grounds that two different phases will be affected differently by a pump, just because they are different.  For example, if a pump corresponds to an increase in temperature, the phase more sensitive to temperature might be more affected; if the pump simply puts energy into a system, the phase with the lower specific heat might be more strongly affected. A pump that directly couples to lattice degrees of freedom would affect a charge density wave phase more strongly than a uniform metallic phase. 
The second most important factor is an asymmetry in the energy landscapes.  If one of the two competing phases has a first order energy landscape (with a metastable zero order parameter state and a globally stable nonzero order parameter state) while the other has a second order energy landscape (zero order parameter state unstable), then the phase with the second order energy landscape is more likely to expand, even if the pump equally heats up the two regions. Differences in relaxation time constants have a more minor effect.   


The considerations of this paper are relevant to recent experiments on strained manganite films \cite{Zhang2016, McLeod2019} where the two competing phases are a charge ordered antiferromagnetic insulator (stable in equilibrium at zero magnetic field) and a  ferromagnetic metal. Moderate fluence optical pulses are found to increase the volume fraction of metastable ferromagnetic metal by moving the interface between metal and insulating phases. This behavior is consistent with the theory presented here because the charge ordering transition is first order whereas in manganites the ferromagnetic tranisition in the absence of charge ordering is second order. Further, empirical evidence suggests that the optical excitation has a more deleterious effect on the charge ordering, because by removing electrons from particular orbitals and by exciting phonons, electronic excitation reduces the the tendency towards  the lattice distortions that are needed for charge ordering.  

One important issue for further research is the explicit inclusion of pinning, which in many situations is necessary to stabilize static interfaces between phases. In this paper we do not explicitly address the pinning issue, focusing instead on the direction that the excited, depinned interface will move. More detailed investigations of the motion of interfaces in the presence of pinning would be desirable, as would extension of the experiments of Refs.~\cite{Zhang2016, McLeod2019} to other systems that also exhibit multiphase coexistence.  The approach presented here may also be relevant to pump-induced phase steering in cuprates \cite{Fausti2011,Cremin2019,Niwa.2019,Zhang2018a}, K$_3$C$_{60}$\cite{Mitrano2016}, FeSe\cite{Suzuki2019a}, SrTiO$_3$\cite{Nova.2019} and other materials\cite{Kogar2019}. Another interesting direction is to consider the effects of non-dissipative dynamics such as those studied in scalar field theory \cite{Boyanovsky.1993} and explicitly in spin models \cite{Turkowski.2006,Lorenzo.2016}.

The linear response result in \equa{eqn:linear_motion}  provides a convenient context to qualitatively discuss the effects of pinning. 
If we take the view that the pump will temporarily depin the interface, the net force on the interface will be the sum of the equilibrium force due to the free energy difference between the two minima, and the transient force applied by the pump.  The interface will move in response to this force and then become pinned again. The equilibrium force may be overcame by the transient force if the pump induced energy asymmetry is larger than the equilibrium one. Of course, in the actual experiments it is likely that fluence beyond the linear response level is required to depin the interface.


\begin{acknowledgements}
We acknowledge support from  the Department of Energy under Grant DE-SC0018218.
\end{acknowledgements}
\bibliographystyle{apsrev4-1}
\bibliography{competing_order_dynamics}

\begin{thebibliography}{27}%
\makeatletter
\providecommand \@ifxundefined [1]{%
 \@ifx{#1\undefined}
}%
\providecommand \@ifnum [1]{%
 \ifnum #1\expandafter \@firstoftwo
 \else \expandafter \@secondoftwo
 \fi
}%
\providecommand \@ifx [1]{%
 \ifx #1\expandafter \@firstoftwo
 \else \expandafter \@secondoftwo
 \fi
}%
\providecommand \natexlab [1]{#1}%
\providecommand \enquote  [1]{``#1''}%
\providecommand \bibnamefont  [1]{#1}%
\providecommand \bibfnamefont [1]{#1}%
\providecommand \citenamefont [1]{#1}%
\providecommand \href@noop [0]{\@secondoftwo}%
\providecommand \href [0]{\begingroup \@sanitize@url \@href}%
\providecommand \@href[1]{\@@startlink{#1}\@@href}%
\providecommand \@@href[1]{\endgroup#1\@@endlink}%
\providecommand \@sanitize@url [0]{\catcode `\\12\catcode `\$12\catcode
  `\&12\catcode `\#12\catcode `\^12\catcode `\_12\catcode `\%12\relax}%
\providecommand \@@startlink[1]{}%
\providecommand \@@endlink[0]{}%
\providecommand \url  [0]{\begingroup\@sanitize@url \@url }%
\providecommand \@url [1]{\endgroup\@href {#1}{\urlprefix }}%
\providecommand \urlprefix  [0]{URL }%
\providecommand \Eprint [0]{\href }%
\providecommand \doibase [0]{http://dx.doi.org/}%
\providecommand \selectlanguage [0]{\@gobble}%
\providecommand \bibinfo  [0]{\@secondoftwo}%
\providecommand \bibfield  [0]{\@secondoftwo}%
\providecommand \translation [1]{[#1]}%
\providecommand \BibitemOpen [0]{}%
\providecommand \bibitemStop [0]{}%
\providecommand \bibitemNoStop [0]{.\EOS\space}%
\providecommand \EOS [0]{\spacefactor3000\relax}%
\providecommand \BibitemShut  [1]{\csname bibitem#1\endcsname}%
\let\auto@bib@innerbib\@empty
\bibitem [{\citenamefont {Domb}\ and\ \citenamefont
  {Lebowitz}(1983)}]{Domb.1983}%
  \BibitemOpen
  \bibinfo {editor} {\bibfnamefont {C.}~\bibnamefont {Domb}}\ and\ \bibinfo
  {editor} {\bibfnamefont {J.~L.}\ \bibnamefont {Lebowitz}},\ eds.,\ \href@noop
  {} {\emph {\bibinfo {title} {Phase Transitions and Critical Phenomena}}},\
  Vol.~\bibinfo {volume} {8}\ (\bibinfo  {publisher} {Academic Press},\
  \bibinfo {address} {Cambridge, Massachusetts},\ \bibinfo {year} {1983})\
  \bibinfo {note} {page 269, Chapter 3}\BibitemShut {NoStop}%
\bibitem [{\citenamefont {Zhang}\ \emph {et~al.}(2016)\citenamefont {Zhang},
  \citenamefont {Tan}, \citenamefont {Liu}, \citenamefont {Teitelbaum},
  \citenamefont {Post}, \citenamefont {Jin}, \citenamefont {Nelson},
  \citenamefont {Basov}, \citenamefont {Wu},\ and\ \citenamefont
  {Averitt}}]{Zhang2016}%
  \BibitemOpen
  \bibfield  {author} {\bibinfo {author} {\bibfnamefont {J.}~\bibnamefont
  {Zhang}}, \bibinfo {author} {\bibfnamefont {X.}~\bibnamefont {Tan}}, \bibinfo
  {author} {\bibfnamefont {M.}~\bibnamefont {Liu}}, \bibinfo {author}
  {\bibfnamefont {S.~W.}\ \bibnamefont {Teitelbaum}}, \bibinfo {author}
  {\bibfnamefont {K.~W.}\ \bibnamefont {Post}}, \bibinfo {author}
  {\bibfnamefont {F.}~\bibnamefont {Jin}}, \bibinfo {author} {\bibfnamefont
  {K.~A.}\ \bibnamefont {Nelson}}, \bibinfo {author} {\bibfnamefont {D.~N.}\
  \bibnamefont {Basov}}, \bibinfo {author} {\bibfnamefont {W.}~\bibnamefont
  {Wu}}, \ and\ \bibinfo {author} {\bibfnamefont {R.~D.}\ \bibnamefont
  {Averitt}},\ }\href {\doibase 10.1038/nmat4695} {\bibfield  {journal}
  {\bibinfo  {journal} {Nat. Mater.}\ }\textbf {\bibinfo {volume} {15}},\
  \bibinfo {pages} {956} (\bibinfo {year} {2016})}\BibitemShut {NoStop}%
\bibitem [{\citenamefont {McLeod}\ \emph {et~al.}(2019)\citenamefont {McLeod},
  \citenamefont {Zhang}, \citenamefont {Gu}, \citenamefont {Jin}, \citenamefont
  {Zhang}, \citenamefont {Post}, \citenamefont {Zhao}, \citenamefont {Millis},
  \citenamefont {Wu}, \citenamefont {Rondinelli}, \citenamefont {Averitt},\
  and\ \citenamefont {Basov}}]{McLeod2019}%
  \BibitemOpen
  \bibfield  {author} {\bibinfo {author} {\bibfnamefont {A.~S.}\ \bibnamefont
  {McLeod}}, \bibinfo {author} {\bibfnamefont {J.}~\bibnamefont {Zhang}},
  \bibinfo {author} {\bibfnamefont {M.~Q.}\ \bibnamefont {Gu}}, \bibinfo
  {author} {\bibfnamefont {F.}~\bibnamefont {Jin}}, \bibinfo {author}
  {\bibfnamefont {G.}~\bibnamefont {Zhang}}, \bibinfo {author} {\bibfnamefont
  {K.~W.}\ \bibnamefont {Post}}, \bibinfo {author} {\bibfnamefont {X.~G.}\
  \bibnamefont {Zhao}}, \bibinfo {author} {\bibfnamefont {A.~J.}\ \bibnamefont
  {Millis}}, \bibinfo {author} {\bibfnamefont {W.~B.}\ \bibnamefont {Wu}},
  \bibinfo {author} {\bibfnamefont {J.~M.}\ \bibnamefont {Rondinelli}},
  \bibinfo {author} {\bibfnamefont {R.~D.}\ \bibnamefont {Averitt}}, \ and\
  \bibinfo {author} {\bibfnamefont {D.~N.}\ \bibnamefont {Basov}},\ }\href
  {https://doi.org/10.1038/s41563-019-0533-y} {\bibfield  {journal} {\bibinfo
  {journal} {Nature Materials}\ ,\ } (\bibinfo {year} {2019})}\BibitemShut
  {NoStop}%
\bibitem [{\citenamefont {Cremin}\ \emph {et~al.}(2019)\citenamefont {Cremin},
  \citenamefont {Zhang}, \citenamefont {Homes}, \citenamefont {Gu},
  \citenamefont {Sun}, \citenamefont {Fogler}, \citenamefont {Millis},
  \citenamefont {Basov},\ and\ \citenamefont {Averitt}}]{Cremin2019}%
  \BibitemOpen
  \bibfield  {author} {\bibinfo {author} {\bibfnamefont {K.~A.}\ \bibnamefont
  {Cremin}}, \bibinfo {author} {\bibfnamefont {J.}~\bibnamefont {Zhang}},
  \bibinfo {author} {\bibfnamefont {C.~C.}\ \bibnamefont {Homes}}, \bibinfo
  {author} {\bibfnamefont {G.~D.}\ \bibnamefont {Gu}}, \bibinfo {author}
  {\bibfnamefont {Z.}~\bibnamefont {Sun}}, \bibinfo {author} {\bibfnamefont
  {M.~M.}\ \bibnamefont {Fogler}}, \bibinfo {author} {\bibfnamefont {A.~J.}\
  \bibnamefont {Millis}}, \bibinfo {author} {\bibfnamefont {D.~N.}\
  \bibnamefont {Basov}}, \ and\ \bibinfo {author} {\bibfnamefont {R.~D.}\
  \bibnamefont {Averitt}},\ }\href {\doibase 10.1073/pnas.1908368116}
  {\bibfield  {journal} {\bibinfo  {journal} {Proceedings of the National
  Academy of Sciences}\ }\textbf {\bibinfo {volume} {116}},\ \bibinfo {pages}
  {19875} (\bibinfo {year} {2019})}\BibitemShut {NoStop}%
\bibitem [{\citenamefont {Fausti}\ \emph {et~al.}(2011)\citenamefont {Fausti},
  \citenamefont {Tobey}, \citenamefont {Dean}, \citenamefont {Kaiser},
  \citenamefont {Dienst}, \citenamefont {Hoffmann}, \citenamefont {Pyon},
  \citenamefont {Takayama}, \citenamefont {Takagi},\ and\ \citenamefont
  {Cavalleri}}]{Fausti2011}%
  \BibitemOpen
  \bibfield  {author} {\bibinfo {author} {\bibfnamefont {D.}~\bibnamefont
  {Fausti}}, \bibinfo {author} {\bibfnamefont {R.~I.}\ \bibnamefont {Tobey}},
  \bibinfo {author} {\bibfnamefont {N.}~\bibnamefont {Dean}}, \bibinfo {author}
  {\bibfnamefont {S.}~\bibnamefont {Kaiser}}, \bibinfo {author} {\bibfnamefont
  {A.}~\bibnamefont {Dienst}}, \bibinfo {author} {\bibfnamefont {M.~C.}\
  \bibnamefont {Hoffmann}}, \bibinfo {author} {\bibfnamefont {S.}~\bibnamefont
  {Pyon}}, \bibinfo {author} {\bibfnamefont {T.}~\bibnamefont {Takayama}},
  \bibinfo {author} {\bibfnamefont {H.}~\bibnamefont {Takagi}}, \ and\ \bibinfo
  {author} {\bibfnamefont {A.}~\bibnamefont {Cavalleri}},\ }\href {\doibase
  10.1126/science.1197294} {\bibfield  {journal} {\bibinfo  {journal}
  {Science}\ }\textbf {\bibinfo {volume} {331}},\ \bibinfo {pages} {189}
  (\bibinfo {year} {2011})}\BibitemShut {NoStop}%
\bibitem [{\citenamefont {Kogar}\ \emph {et~al.}(2019)\citenamefont {Kogar},
  \citenamefont {Zong}, \citenamefont {Dolgirev}, \citenamefont {Shen},
  \citenamefont {Straquadine}, \citenamefont {Bie}, \citenamefont {Wang},
  \citenamefont {Rohwer}, \citenamefont {Tung}, \citenamefont {Yang},
  \citenamefont {Li}, \citenamefont {Yang}, \citenamefont {Weathersby},
  \citenamefont {Park}, \citenamefont {Kozina}, \citenamefont {Sie},
  \citenamefont {Wen}, \citenamefont {Jarillo-Herrero}, \citenamefont {Fisher},
  \citenamefont {Wang},\ and\ \citenamefont {Gedik}}]{Kogar2019}%
  \BibitemOpen
  \bibfield  {author} {\bibinfo {author} {\bibfnamefont {A.}~\bibnamefont
  {Kogar}}, \bibinfo {author} {\bibfnamefont {A.}~\bibnamefont {Zong}},
  \bibinfo {author} {\bibfnamefont {P.~E.}\ \bibnamefont {Dolgirev}}, \bibinfo
  {author} {\bibfnamefont {X.}~\bibnamefont {Shen}}, \bibinfo {author}
  {\bibfnamefont {J.}~\bibnamefont {Straquadine}}, \bibinfo {author}
  {\bibfnamefont {Y.-Q.}\ \bibnamefont {Bie}}, \bibinfo {author} {\bibfnamefont
  {X.}~\bibnamefont {Wang}}, \bibinfo {author} {\bibfnamefont {T.}~\bibnamefont
  {Rohwer}}, \bibinfo {author} {\bibfnamefont {I.-C.}\ \bibnamefont {Tung}},
  \bibinfo {author} {\bibfnamefont {Y.}~\bibnamefont {Yang}}, \bibinfo {author}
  {\bibfnamefont {R.}~\bibnamefont {Li}}, \bibinfo {author} {\bibfnamefont
  {J.}~\bibnamefont {Yang}}, \bibinfo {author} {\bibfnamefont {S.}~\bibnamefont
  {Weathersby}}, \bibinfo {author} {\bibfnamefont {S.}~\bibnamefont {Park}},
  \bibinfo {author} {\bibfnamefont {M.~E.}\ \bibnamefont {Kozina}}, \bibinfo
  {author} {\bibfnamefont {E.~J.}\ \bibnamefont {Sie}}, \bibinfo {author}
  {\bibfnamefont {H.}~\bibnamefont {Wen}}, \bibinfo {author} {\bibfnamefont
  {P.}~\bibnamefont {Jarillo-Herrero}}, \bibinfo {author} {\bibfnamefont
  {I.~R.}\ \bibnamefont {Fisher}}, \bibinfo {author} {\bibfnamefont
  {X.}~\bibnamefont {Wang}}, \ and\ \bibinfo {author} {\bibfnamefont
  {N.}~\bibnamefont {Gedik}},\ }\href
  {https://doi.org/10.1038/s41567-019-0705-3} {\bibfield  {journal} {\bibinfo
  {journal} {Nature Physics}\ ,\ } (\bibinfo {year} {2019})}\BibitemShut
  {NoStop}%
\bibitem [{\citenamefont {Sun}\ and\ \citenamefont {Millis}(2020)}]{Sun2019}%
  \BibitemOpen
  \bibfield  {author} {\bibinfo {author} {\bibfnamefont {Z.}~\bibnamefont
  {Sun}}\ and\ \bibinfo {author} {\bibfnamefont {A.~J.}\ \bibnamefont
  {Millis}},\ }\href {\doibase 10.1103/PhysRevX.10.021028} {\bibfield
  {journal} {\bibinfo  {journal} {Phys. Rev. X}\ }\textbf {\bibinfo {volume}
  {10}},\ \bibinfo {pages} {021028} (\bibinfo {year} {2020})}\BibitemShut
  {NoStop}%
\bibitem [{\citenamefont {Zhang}\ \emph
  {et~al.}(2018{\natexlab{a}})\citenamefont {Zhang}, \citenamefont {Wang},
  \citenamefont {Wu}, \citenamefont {Liu}, \citenamefont {Shi}, \citenamefont
  {Lin}, \citenamefont {Li}, \citenamefont {Dai}, \citenamefont {Dong},\ and\
  \citenamefont {Wang}}]{Zhang2018}%
  \BibitemOpen
  \bibfield  {author} {\bibinfo {author} {\bibfnamefont {S.~J.}\ \bibnamefont
  {Zhang}}, \bibinfo {author} {\bibfnamefont {Z.~X.}\ \bibnamefont {Wang}},
  \bibinfo {author} {\bibfnamefont {D.}~\bibnamefont {Wu}}, \bibinfo {author}
  {\bibfnamefont {Q.~M.}\ \bibnamefont {Liu}}, \bibinfo {author} {\bibfnamefont
  {L.~Y.}\ \bibnamefont {Shi}}, \bibinfo {author} {\bibfnamefont
  {T.}~\bibnamefont {Lin}}, \bibinfo {author} {\bibfnamefont {S.~L.}\
  \bibnamefont {Li}}, \bibinfo {author} {\bibfnamefont {P.~C.}\ \bibnamefont
  {Dai}}, \bibinfo {author} {\bibfnamefont {T.}~\bibnamefont {Dong}}, \ and\
  \bibinfo {author} {\bibfnamefont {N.~L.}\ \bibnamefont {Wang}},\ }\href
  {\doibase 10.1103/PhysRevB.98.224507} {\bibfield  {journal} {\bibinfo
  {journal} {Phys. Rev. B}\ }\textbf {\bibinfo {volume} {98}},\ \bibinfo
  {pages} {224507} (\bibinfo {year} {2018}{\natexlab{a}})}\BibitemShut
  {NoStop}%
\bibitem [{\citenamefont {Gerasimenko}\ \emph {et~al.}(2019)\citenamefont
  {Gerasimenko}, \citenamefont {Karpov}, \citenamefont {Vaskivskyi},
  \citenamefont {Brazovskii},\ and\ \citenamefont
  {Mihailovic}}]{Gerasimenko2019}%
  \BibitemOpen
  \bibfield  {author} {\bibinfo {author} {\bibfnamefont {Y.~A.}\ \bibnamefont
  {Gerasimenko}}, \bibinfo {author} {\bibfnamefont {P.}~\bibnamefont {Karpov}},
  \bibinfo {author} {\bibfnamefont {I.}~\bibnamefont {Vaskivskyi}}, \bibinfo
  {author} {\bibfnamefont {S.}~\bibnamefont {Brazovskii}}, \ and\ \bibinfo
  {author} {\bibfnamefont {D.}~\bibnamefont {Mihailovic}},\ }\href
  {https://doi.org/10.1038/s41535-019-0172-1} {\bibfield  {journal} {\bibinfo
  {journal} {npj Quantum Materials}\ }\textbf {\bibinfo {volume} {4}},\
  \bibinfo {pages} {32} (\bibinfo {year} {2019})}\BibitemShut {NoStop}%
\bibitem [{\citenamefont {Nova}\ \emph {et~al.}(2019)\citenamefont {Nova},
  \citenamefont {Disa}, \citenamefont {Fechner},\ and\ \citenamefont
  {Cavalleri}}]{Nova.2019}%
  \BibitemOpen
  \bibfield  {author} {\bibinfo {author} {\bibfnamefont {T.~F.}\ \bibnamefont
  {Nova}}, \bibinfo {author} {\bibfnamefont {A.~S.}\ \bibnamefont {Disa}},
  \bibinfo {author} {\bibfnamefont {M.}~\bibnamefont {Fechner}}, \ and\
  \bibinfo {author} {\bibfnamefont {A.}~\bibnamefont {Cavalleri}},\ }\href
  {\doibase 10.1126/science.aaw4911} {\bibfield  {journal} {\bibinfo  {journal}
  {Science}\ }\textbf {\bibinfo {volume} {364}},\ \bibinfo {pages} {1075}
  (\bibinfo {year} {2019})}\BibitemShut {NoStop}%
\bibitem [{\citenamefont {Kung}\ \emph {et~al.}(2013)\citenamefont {Kung},
  \citenamefont {Lee}, \citenamefont {Chen}, \citenamefont {Kemper},
  \citenamefont {Sorini}, \citenamefont {Moritz},\ and\ \citenamefont
  {Devereaux}}]{Kung2013}%
  \BibitemOpen
  \bibfield  {author} {\bibinfo {author} {\bibfnamefont {Y.~F.}\ \bibnamefont
  {Kung}}, \bibinfo {author} {\bibfnamefont {W.-S.}\ \bibnamefont {Lee}},
  \bibinfo {author} {\bibfnamefont {C.-C.}\ \bibnamefont {Chen}}, \bibinfo
  {author} {\bibfnamefont {A.~F.}\ \bibnamefont {Kemper}}, \bibinfo {author}
  {\bibfnamefont {A.~P.}\ \bibnamefont {Sorini}}, \bibinfo {author}
  {\bibfnamefont {B.}~\bibnamefont {Moritz}}, \ and\ \bibinfo {author}
  {\bibfnamefont {T.~P.}\ \bibnamefont {Devereaux}},\ }\href {\doibase
  10.1103/PhysRevB.88.125114} {\bibfield  {journal} {\bibinfo  {journal} {Phys.
  Rev. B}\ }\textbf {\bibinfo {volume} {88}},\ \bibinfo {pages} {125114}
  (\bibinfo {year} {2013})}\BibitemShut {NoStop}%
\bibitem [{\citenamefont {Ross~Tagaras}\ \emph {et~al.}(2019)\citenamefont
  {Ross~Tagaras}, \citenamefont {Weng},\ and\ \citenamefont
  {Allen}}]{RossTagaras2019}%
  \BibitemOpen
  \bibfield  {author} {\bibinfo {author} {\bibfnamefont {M.}~\bibnamefont
  {Ross~Tagaras}}, \bibinfo {author} {\bibfnamefont {J.}~\bibnamefont {Weng}},
  \ and\ \bibinfo {author} {\bibfnamefont {R.~E.}\ \bibnamefont {Allen}},\
  }\href {https://doi.org/10.1140/epjst/e2018-800102-6} {\bibfield  {journal}
  {\bibinfo  {journal} {The European Physical Journal Special Topics}\ }\textbf
  {\bibinfo {volume} {227}},\ \bibinfo {pages} {2297} (\bibinfo {year}
  {2019})}\BibitemShut {NoStop}%
\bibitem [{\citenamefont {Dolgirev}\ \emph {et~al.}(2020)\citenamefont
  {Dolgirev}, \citenamefont {Rozhkov}, \citenamefont {Zong}, \citenamefont
  {Kogar}, \citenamefont {Gedik},\ and\ \citenamefont {Fine}}]{Dolgirev2019}%
  \BibitemOpen
  \bibfield  {author} {\bibinfo {author} {\bibfnamefont {P.~E.}\ \bibnamefont
  {Dolgirev}}, \bibinfo {author} {\bibfnamefont {A.~V.}\ \bibnamefont
  {Rozhkov}}, \bibinfo {author} {\bibfnamefont {A.}~\bibnamefont {Zong}},
  \bibinfo {author} {\bibfnamefont {A.}~\bibnamefont {Kogar}}, \bibinfo
  {author} {\bibfnamefont {N.}~\bibnamefont {Gedik}}, \ and\ \bibinfo {author}
  {\bibfnamefont {B.~V.}\ \bibnamefont {Fine}},\ }\href {\doibase
  10.1103/PhysRevB.101.054203} {\bibfield  {journal} {\bibinfo  {journal}
  {Phys. Rev. B}\ }\textbf {\bibinfo {volume} {101}},\ \bibinfo {pages}
  {054203} (\bibinfo {year} {2020})}\BibitemShut {NoStop}%
\bibitem [{\citenamefont {Tokura}(2006)}]{Tokura06}%
  \BibitemOpen
  \bibfield  {author} {\bibinfo {author} {\bibfnamefont {Y.}~\bibnamefont
  {Tokura}},\ }\href {\doibase 10.1088/0034-4885/69/3/r06} {\bibfield
  {journal} {\bibinfo  {journal} {Reports on Progress in Physics}\ }\textbf
  {\bibinfo {volume} {69}},\ \bibinfo {pages} {797} (\bibinfo {year}
  {2006})}\BibitemShut {NoStop}%
\bibitem [{\citenamefont {Seman}\ \emph {et~al.}(2012)\citenamefont {Seman},
  \citenamefont {Ahn}, \citenamefont {Lookman}, \citenamefont {Saxena},
  \citenamefont {Bishop},\ and\ \citenamefont {Littlewood}}]{Seman12}%
  \BibitemOpen
  \bibfield  {author} {\bibinfo {author} {\bibfnamefont {T.~F.}\ \bibnamefont
  {Seman}}, \bibinfo {author} {\bibfnamefont {K.~H.}\ \bibnamefont {Ahn}},
  \bibinfo {author} {\bibfnamefont {T.}~\bibnamefont {Lookman}}, \bibinfo
  {author} {\bibfnamefont {A.}~\bibnamefont {Saxena}}, \bibinfo {author}
  {\bibfnamefont {A.~R.}\ \bibnamefont {Bishop}}, \ and\ \bibinfo {author}
  {\bibfnamefont {P.~B.}\ \bibnamefont {Littlewood}},\ }\href {\doibase
  10.1103/PhysRevB.86.184106} {\bibfield  {journal} {\bibinfo  {journal} {Phys.
  Rev. B}\ }\textbf {\bibinfo {volume} {86}},\ \bibinfo {pages} {184106}
  (\bibinfo {year} {2012})}\BibitemShut {NoStop}%
\bibitem [{\citenamefont {Gor'kov}\ and\ \citenamefont
  {Eliashberg}(1968)}]{Gorkov1968}%
  \BibitemOpen
  \bibfield  {author} {\bibinfo {author} {\bibfnamefont {L.~P.}\ \bibnamefont
  {Gor'kov}}\ and\ \bibinfo {author} {\bibfnamefont {G.~M.}\ \bibnamefont
  {Eliashberg}},\ }\href@noop {} {\bibfield  {journal} {\bibinfo  {journal}
  {Sov. Phys. JETP}\ }\textbf {\bibinfo {volume} {27}},\ \bibinfo {pages} {328}
  (\bibinfo {year} {1968})}\BibitemShut {NoStop}%
\bibitem [{\citenamefont {Cyrot}(1973)}]{Cyrot1973}%
  \BibitemOpen
  \bibfield  {author} {\bibinfo {author} {\bibfnamefont {M.}~\bibnamefont
  {Cyrot}},\ }\href {\doibase 10.1088/0034-4885/36/2/001} {\bibfield  {journal}
  {\bibinfo  {journal} {Reports Prog. Phys.}\ }\textbf {\bibinfo {volume}
  {36}},\ \bibinfo {pages} {103} (\bibinfo {year} {1973})}\BibitemShut
  {NoStop}%
\bibitem [{\citenamefont {Hohenberg}\ and\ \citenamefont
  {Halperin}(1977)}]{Hohenberg1977}%
  \BibitemOpen
  \bibfield  {author} {\bibinfo {author} {\bibfnamefont {P.~C.}\ \bibnamefont
  {Hohenberg}}\ and\ \bibinfo {author} {\bibfnamefont {B.~I.}\ \bibnamefont
  {Halperin}},\ }\href {\doibase 10.1103/RevModPhys.49.435} {\bibfield
  {journal} {\bibinfo  {journal} {Rev. Mod. Phys.}\ }\textbf {\bibinfo {volume}
  {49}},\ \bibinfo {pages} {435} (\bibinfo {year} {1977})}\BibitemShut
  {NoStop}%
\bibitem [{\citenamefont {Lemonik}\ and\ \citenamefont
  {Mitra}(2017)}]{Lemonik.2017}%
  \BibitemOpen
  \bibfield  {author} {\bibinfo {author} {\bibfnamefont {Y.}~\bibnamefont
  {Lemonik}}\ and\ \bibinfo {author} {\bibfnamefont {A.}~\bibnamefont
  {Mitra}},\ }\href {\doibase 10.1103/PhysRevB.96.104506} {\bibfield  {journal}
  {\bibinfo  {journal} {Phys. Rev. B}\ }\textbf {\bibinfo {volume} {96}},\
  \bibinfo {pages} {104506} (\bibinfo {year} {2017})}\BibitemShut {NoStop}%
\bibitem [{\citenamefont {Sun}\ \emph {et~al.}(2020)\citenamefont {Sun},
  \citenamefont {Fogler}, \citenamefont {Basov},\ and\ \citenamefont
  {Millis}}]{sun2020collective}%
  \BibitemOpen
  \bibfield  {author} {\bibinfo {author} {\bibfnamefont {Z.}~\bibnamefont
  {Sun}}, \bibinfo {author} {\bibfnamefont {M.~M.}\ \bibnamefont {Fogler}},
  \bibinfo {author} {\bibfnamefont {D.~N.}\ \bibnamefont {Basov}}, \ and\
  \bibinfo {author} {\bibfnamefont {A.~J.}\ \bibnamefont {Millis}},\
  }\href@noop {} {\enquote {\bibinfo {title} {Collective modes and thz near
  field response of superconductors},}\ } (\bibinfo {year} {2020}),\ \Eprint
  {http://arxiv.org/abs/2001.03704} {arXiv:2001.03704 [cond-mat.supr-con]}
  \BibitemShut {NoStop}%
\bibitem [{\citenamefont {Niwa}\ \emph {et~al.}(2019)\citenamefont {Niwa},
  \citenamefont {Yoshikawa}, \citenamefont {Tomari}, \citenamefont {Matsunaga},
  \citenamefont {Song}, \citenamefont {Eisaki},\ and\ \citenamefont
  {Shimano}}]{Niwa.2019}%
  \BibitemOpen
  \bibfield  {author} {\bibinfo {author} {\bibfnamefont {H.}~\bibnamefont
  {Niwa}}, \bibinfo {author} {\bibfnamefont {N.}~\bibnamefont {Yoshikawa}},
  \bibinfo {author} {\bibfnamefont {K.}~\bibnamefont {Tomari}}, \bibinfo
  {author} {\bibfnamefont {R.}~\bibnamefont {Matsunaga}}, \bibinfo {author}
  {\bibfnamefont {D.}~\bibnamefont {Song}}, \bibinfo {author} {\bibfnamefont
  {H.}~\bibnamefont {Eisaki}}, \ and\ \bibinfo {author} {\bibfnamefont
  {R.}~\bibnamefont {Shimano}},\ }\href {\doibase 10.1103/PhysRevB.100.104507}
  {\bibfield  {journal} {\bibinfo  {journal} {Phys. Rev. B}\ }\textbf {\bibinfo
  {volume} {100}},\ \bibinfo {pages} {104507} (\bibinfo {year}
  {2019})}\BibitemShut {NoStop}%
\bibitem [{\citenamefont {Zhang}\ \emph
  {et~al.}(2018{\natexlab{b}})\citenamefont {Zhang}, \citenamefont {Wang},
  \citenamefont {Shi}, \citenamefont {Lin}, \citenamefont {Zhang},
  \citenamefont {Gu}, \citenamefont {Dong},\ and\ \citenamefont
  {Wang}}]{Zhang2018a}%
  \BibitemOpen
  \bibfield  {author} {\bibinfo {author} {\bibfnamefont {S.~J.}\ \bibnamefont
  {Zhang}}, \bibinfo {author} {\bibfnamefont {Z.~X.}\ \bibnamefont {Wang}},
  \bibinfo {author} {\bibfnamefont {L.~Y.}\ \bibnamefont {Shi}}, \bibinfo
  {author} {\bibfnamefont {T.}~\bibnamefont {Lin}}, \bibinfo {author}
  {\bibfnamefont {M.~Y.}\ \bibnamefont {Zhang}}, \bibinfo {author}
  {\bibfnamefont {G.~D.}\ \bibnamefont {Gu}}, \bibinfo {author} {\bibfnamefont
  {T.}~\bibnamefont {Dong}}, \ and\ \bibinfo {author} {\bibfnamefont {N.~L.}\
  \bibnamefont {Wang}},\ }\href {\doibase 10.1103/PhysRevB.98.020506}
  {\bibfield  {journal} {\bibinfo  {journal} {Phys. Rev. B}\ }\textbf {\bibinfo
  {volume} {98}},\ \bibinfo {pages} {020506(R)} (\bibinfo {year}
  {2018}{\natexlab{b}})}\BibitemShut {NoStop}%
\bibitem [{\citenamefont {Mitrano}\ \emph {et~al.}(2016)\citenamefont
  {Mitrano}, \citenamefont {Cantaluppi}, \citenamefont {Nicoletti},
  \citenamefont {Kaiser}, \citenamefont {Perucchi}, \citenamefont {Lupi},
  \citenamefont {{Di Pietro}}, \citenamefont {Pontiroli}, \citenamefont
  {Ricc{\`{o}}}, \citenamefont {Clark}, \citenamefont {Jaksch},\ and\
  \citenamefont {Cavalleri}}]{Mitrano2016}%
  \BibitemOpen
  \bibfield  {author} {\bibinfo {author} {\bibfnamefont {M.}~\bibnamefont
  {Mitrano}}, \bibinfo {author} {\bibfnamefont {A.}~\bibnamefont {Cantaluppi}},
  \bibinfo {author} {\bibfnamefont {D.}~\bibnamefont {Nicoletti}}, \bibinfo
  {author} {\bibfnamefont {S.}~\bibnamefont {Kaiser}}, \bibinfo {author}
  {\bibfnamefont {A.}~\bibnamefont {Perucchi}}, \bibinfo {author}
  {\bibfnamefont {S.}~\bibnamefont {Lupi}}, \bibinfo {author} {\bibfnamefont
  {P.}~\bibnamefont {{Di Pietro}}}, \bibinfo {author} {\bibfnamefont
  {D.}~\bibnamefont {Pontiroli}}, \bibinfo {author} {\bibfnamefont
  {M.}~\bibnamefont {Ricc{\`{o}}}}, \bibinfo {author} {\bibfnamefont {S.~R.}\
  \bibnamefont {Clark}}, \bibinfo {author} {\bibfnamefont {D.}~\bibnamefont
  {Jaksch}}, \ and\ \bibinfo {author} {\bibfnamefont {A.}~\bibnamefont
  {Cavalleri}},\ }\href {\doibase 10.1038/nature16522} {\bibfield  {journal}
  {\bibinfo  {journal} {Nature}\ }\textbf {\bibinfo {volume} {530}},\ \bibinfo
  {pages} {461} (\bibinfo {year} {2016})}\BibitemShut {NoStop}%
\bibitem [{\citenamefont {Suzuki}\ \emph {et~al.}(2019)\citenamefont {Suzuki},
  \citenamefont {Someya}, \citenamefont {Hashimoto}, \citenamefont {Michimae},
  \citenamefont {Watanabe}, \citenamefont {Fujisawa}, \citenamefont {Kanai},
  \citenamefont {Ishii}, \citenamefont {Itatani}, \citenamefont {Kasahara},
  \citenamefont {Matsuda}, \citenamefont {Shibauchi}, \citenamefont {Okazaki},\
  and\ \citenamefont {Shin}}]{Suzuki2019a}%
  \BibitemOpen
  \bibfield  {author} {\bibinfo {author} {\bibfnamefont {T.}~\bibnamefont
  {Suzuki}}, \bibinfo {author} {\bibfnamefont {T.}~\bibnamefont {Someya}},
  \bibinfo {author} {\bibfnamefont {T.}~\bibnamefont {Hashimoto}}, \bibinfo
  {author} {\bibfnamefont {S.}~\bibnamefont {Michimae}}, \bibinfo {author}
  {\bibfnamefont {M.}~\bibnamefont {Watanabe}}, \bibinfo {author}
  {\bibfnamefont {M.}~\bibnamefont {Fujisawa}}, \bibinfo {author}
  {\bibfnamefont {T.}~\bibnamefont {Kanai}}, \bibinfo {author} {\bibfnamefont
  {N.}~\bibnamefont {Ishii}}, \bibinfo {author} {\bibfnamefont
  {J.}~\bibnamefont {Itatani}}, \bibinfo {author} {\bibfnamefont
  {S.}~\bibnamefont {Kasahara}}, \bibinfo {author} {\bibfnamefont
  {Y.}~\bibnamefont {Matsuda}}, \bibinfo {author} {\bibfnamefont
  {T.}~\bibnamefont {Shibauchi}}, \bibinfo {author} {\bibfnamefont
  {K.}~\bibnamefont {Okazaki}}, \ and\ \bibinfo {author} {\bibfnamefont
  {S.}~\bibnamefont {Shin}},\ }\href
  {https://doi.org/10.1038/s42005-019-0219-4} {\bibfield  {journal} {\bibinfo
  {journal} {Communications Physics}\ }\textbf {\bibinfo {volume} {2}},\
  \bibinfo {pages} {115} (\bibinfo {year} {2019})}\BibitemShut {NoStop}%
\bibitem [{\citenamefont {Boyanovsky}(1993)}]{Boyanovsky.1993}%
  \BibitemOpen
  \bibfield  {author} {\bibinfo {author} {\bibfnamefont {D.}~\bibnamefont
  {Boyanovsky}},\ }\href {\doibase 10.1103/PhysRevE.48.767} {\bibfield
  {journal} {\bibinfo  {journal} {Phys. Rev. E}\ }\textbf {\bibinfo {volume}
  {48}},\ \bibinfo {pages} {767} (\bibinfo {year} {1993})}\BibitemShut
  {NoStop}%
\bibitem [{\citenamefont {Turkowski}\ \emph {et~al.}(2006)\citenamefont
  {Turkowski}, \citenamefont {Sacramento},\ and\ \citenamefont
  {Vieira}}]{Turkowski.2006}%
  \BibitemOpen
  \bibfield  {author} {\bibinfo {author} {\bibfnamefont {V.~M.}\ \bibnamefont
  {Turkowski}}, \bibinfo {author} {\bibfnamefont {P.~D.}\ \bibnamefont
  {Sacramento}}, \ and\ \bibinfo {author} {\bibfnamefont {V.~R.}\ \bibnamefont
  {Vieira}},\ }\href {\doibase 10.1103/PhysRevB.73.214437} {\bibfield
  {journal} {\bibinfo  {journal} {Phys. Rev. B}\ }\textbf {\bibinfo {volume}
  {73}},\ \bibinfo {pages} {214437} (\bibinfo {year} {2006})}\BibitemShut
  {NoStop}%
\bibitem [{\citenamefont {{Del Re}}\ \emph {et~al.}(2016)\citenamefont {{Del
  Re}}, \citenamefont {Fabrizio},\ and\ \citenamefont
  {Tosatti}}]{Lorenzo.2016}%
  \BibitemOpen
  \bibfield  {author} {\bibinfo {author} {\bibfnamefont {L.}~\bibnamefont {{Del
  Re}}}, \bibinfo {author} {\bibfnamefont {M.}~\bibnamefont {Fabrizio}}, \ and\
  \bibinfo {author} {\bibfnamefont {E.}~\bibnamefont {Tosatti}},\ }\href
  {\doibase 10.1103/PhysRevB.93.125131} {\bibfield  {journal} {\bibinfo
  {journal} {Phys. Rev. B}\ }\textbf {\bibinfo {volume} {93}},\ \bibinfo
  {pages} {125131} (\bibinfo {year} {2016})}\BibitemShut {NoStop}%
\end{thebibliography}%

\end{document}